\documentclass[twocolumn,aps,prd,showpacs,nofootinbib,floatfix,eqsecnum]{revtex4}
\usepackage{amsmath,latexsym,graphicx,multirow,longtable} 
\begin{document}
\title{The transition from adiabatic inspiral to geodesic plunge for a compact object around a massive Kerr black hole: Generic orbits}
\author{Pranesh A Sundararajan}
\affiliation{Department of Physics and MIT Kavli Institute, MIT, 77
Massachusetts Ave., Cambridge, MA 02139}
\date{\today}
\begin{abstract}
The inspiral of a stellar mass compact object falling into a massive Kerr black hole can be broken into three different regimes:  An adiabatic inspiral phase, where the inspiral timescale is much larger than the orbital period; a late-time radial infall, which can be approximated as a plunging geodesic; and a regime where the body transitions from the inspiral to plunge.  In earlier work, Ori and Thorne have outlined a method to compute the trajectory during this  transition for a compact object in a circular, equatorial orbit. We generalize this technique to include inclination and eccentricity.  
\end{abstract}
\maketitle

\section{Introduction and motivation}
Extreme mass ratio inspirals (EMRIs), in which stellar mass compact objects radiate gravitational energy and fall into their massive black hole companions, are promising sources of gravitational waves. LISA \cite{lisa}, the proposed space based gravitational wave detector should detect waves from the last stages of such inspirals. A clear theoretical understanding of the dynamics of EMRIs is vital to the detection of these gravitational waves.

The small mass ratios, which typically lie in the range $\mu/M = 10^{-5}-10^{-8}$, allow 
EMRIs to be treated within the framework of perturbation theory. The trajectory of the compact object can be roughly broken  into three regimes: (a) An adiabatic inspiral phase, during which the dominant inspiral mechanism arises from the radiation reaction force on the smaller  object. In this stage, the time scale over which the characteristic radial separation (between the compact object and its black hole companion) changes is large compared to the orbital period. This allows us to approximate the trajectory as a sequence of bound geodesics. (b) A plunge phase, during which stable geodesics do not exist. It has been shown \cite{ot2000} that the effect of radiation reaction is negligible during the plunge and that this phase can be modeled as a geodesic infall. (c) A  regime where the spiraling compact object transitions from  adiabatic inspiral to  geodesic plunge. The course of motion at this juncture shows aspects of both, the self-force from radiation reaction and the effects of unstable geodesics. 

In \cite{ot2000}, Ori and Thorne introduce a  method to predict the motion  when the object is constrained to an approximately circular, equatorial orbit. We generalize this procedure  to include inclined and eccentric trajectories. A few modifications to the prescription in \cite{ot2000} are introduced to handle such generic orbits. The results from our generalized prescription are in excellent agreement with \cite{ot2000}. 

The simple calculation described in this paper is meant to serve as a stopgap for many other open and important problems. There has been recent progress (\cite{skh07, skhd08} and references therein) in the development of a code to solve the Teukolsky equation in the time-domain. The world line of the compact object serves as an input to this code. While the world line in the adiabatic phase can be calculated from a frequency-domain based Teukolsky equation solver \cite{hughes2000, dh}, the trajectory in the transition regime for completely generic obits remains unknown. This calculation will provide the missing link needed to generate a complete inspiral trajectory. 

A number of researchers are working towards solving the self-force problem exactly \cite{sf1,sf2}. Such an exact solution can be separated (at least qualitatively) into time-reversal symmetric and asymmetric components. The symmetric component (the ``conservative self-force'') conserves the integrals of motion. On the other hand, the asymmetric component (the ``dissipative self-force'')  leads to non-zero time derivatives of the integrals of motion.  Recent advances demonstrate that we are making steady progress on this problem.
For example, the self force is now essentially understood for circular orbits
around Schwarzschild black holes \cite{sf2}. Although approximate, the results in this paper may serve as an independent check for these solutions. It is worth noting that if it becomes possible to include the conservative force in a simple way, we should be able to build its impact into the formalism developed here.  
This work may also be of interest for numerical relativity --- a perturbative inspiral
constructed by the techniques discussed here may be an accurate point of comparison for
full numerical inspirals for small ratios (and may even be useful, if not so accurate, for
mass ratios that are not strictly perturbative).

Ref.\ \cite{eq_tr} discusses the transition when the compact object is in an eccentric, equatorial orbit. However, the focus of that paper is to calculate the transit time and estimate the probability for LISA to observe such a transition. Our intent is to generate the world line during the transition.  We also choose our initial conditions
differently than they are chosen in Ref.\ \cite{eq_tr}; we discuss these differences in more detail in Sec.\ \ref{sec:ecc}.  

The rest of the paper is organized as follows: Section \ref{sec:circ} discusses circular orbits with arbitrary inclination. Sec. \ref{sec:ecc} generalizes the formalism developed in Sec.\ \ref{sec:circ} to include eccentricity. Finally, we summarize our results in Sec.\ \ref{sec:summ}.

\section{The transition trajectory for circular orbits}
\label{sec:circ}
Up to initial conditions, a set of three constants, the energy, $E$, the component of the angular momentum along the spin axis, $L_z$, and the Carter constant, $Q$ define a geodesic. The Carter constant has an approximate interpretation of being the square of the component of angular momentum perpendicular to the spin axis. As the compact object radiates, the ``constants'' that define its geodesic will gradually evolve. (We will refer to $[E(t),L_z(t),Q(t)]$ as the ``constants'', although they are slowly evolving.) A common approach to model the adiabatic regime consists of treating the motion as the sequence of geodesics \cite{hughes2000,dh} defined by these evolving constants. As pointed out in \cite{pp}, this limit amounts to a ``radiative'' or ``dissipative'' approximation.  A true adiabatic approximation would be a sequence of orbits in which each orbit included conservative self corrections.  Since we currently use purely geodesic orbits as
our background motion (in lieu of a self-force enhanced description), we will refer to a
sequence of geodesics as an ``adiabatic inspiral" throughout this paper. Thus, within the adiabatic approximation, the world line of a particle is computed by mapping $[E(t), L_z(t), Q(t)]$ to $[r(t),\theta(t),\phi(t)]$. The symbols $r$, $\theta$ and $\phi$ are the usual Boyer-Lindquist coordinates.

In contrast, the plunge can be treated as a single unstable geodesic with almost constant $E$, $L_z$ and $Q$. Thus, the passage from adiabatic inspiral to geodesic plunge must contain both these features --- slowly evolving ``constants'' and marginal stability.
 
\subsection{Kerr Geodesics}
The following system of first order equations describes geodesics in a Kerr \cite{kerr1, kerr2} geometry:  
\begin{eqnarray}
\label{geod1}
 \Sigma \frac{dr}{d\tau} = \pm \sqrt{R} \; , \\
\label{geod2} 
\Sigma \frac{d\theta}{d\tau} = \pm \sqrt{V_\theta} \; , \\
\label{geod3} 
\Sigma \frac{d\phi}{d\tau} = V_\phi \; , \\
\label{geod4} 
\Sigma \frac{dt}{d\tau} = V_t \;.
\end{eqnarray}
The potentials can be expressed as:
\begin{eqnarray}
\label{eq:R}
R & = &   \frac{1}{\mu^2}\left[E \left(a^2+r^2\right)-a L_z\right]^2-  \nonumber \\
	&  &  +\frac{\Delta}{\mu^2}  \left[(L_z-a E)^2+r^2 \mu ^2+Q\right] \;, \\
V_\theta & = &  \frac{1}{\mu^2}\left[Q - \right. \nonumber \\
	& & \left. \cos^2\theta\left(a^2\left(\mu^2-E^2\right) + L_z^2/\sin^2\theta\right)\right] \;,\\
V_\phi & = & \frac{1}{\mu}\left[L_z/\sin^2\theta - a E \right] \nonumber \\
	&  & + \frac{a}{\mu\Delta}\left[E\left(r^2 + a^2\right)-L_z a\right] \;, \\
\label{eq:Vt}
V_t & = & \frac{1}{\mu} \left[a\left(L_z - a E \sin^2\theta\right) \right] + \nonumber \\
    &  & \frac{r^2 + a^2}{\mu \Delta}\left[E\left(r^2 + a^2\right) - L_z a\right] \;.
\end{eqnarray}
The parameters $(r, \theta, \phi, t)$ are the Boyer-Lindquist coordinates, $M$ is the black hole mass, $\mu$ is the perturbing mass, $\Sigma = r^2 + a^2 \cos^2\theta$, $\Delta = r^2 - 2 M r + a^2$ and $a$ is the spin parameter of the black hole. The constants $(E, L_z, Q)$ represent the {\it{actual}} energy, momentum and Carter constant (in units of $M$, $M^2$ and $M^4$ respectively), not the dimensionless versions of them. By introducing the perturbing mass explicitly, our notation deviates from previous literature. We do this in order to show the dependence of the transition phase on the mass of the perturbing object. We also set $G=c=1$ everywhere.

\subsection{The last stable orbit}
A standard but not unique definition of the ``inclination'' of a Kerr geodesic is given by
\begin{eqnarray}
\label{eq:Q}
\cos\iota & = & \frac{L_z}{\sqrt{L_z^2 + Q}}\;, \\
\Rightarrow Q & = & \frac{L_z^2}{\cos^2\iota - L_z^2} \;.
\end{eqnarray}
It is possible to use $\iota$ to eliminate the Carter constant. Thus, any circular orbit can be parametrized by its radius ($r$) and inclination ($\iota$). 

The last stable orbit (LSO) serves as an important reference point --- the inspiral is adiabatic well before the compact object crosses the LSO and is approximately a plunge well after the crossing. Since the transition occurs in the vicinity of the LSO, a preliminary step in our computation is to determine $r$ and $(E,L_z,Q)$ at the LSO for a given inclination at the LSO, $\iota_{\rm LSO}$. Note that $\iota$ changes with time because it is a function of $[E(t),L_z(t),Q(t)]$. 

Circular orbits satisfy 
\begin{eqnarray}
R & = &0 \; \mbox{and} \\
R^\prime & = & \frac{dR}{dr} = 0 \;.
\end{eqnarray}
We must have $R^\prime = 0$ because the LSO lies at an extremum of $R$. We also require that 
\begin{eqnarray}
R^{\prime\prime} & = & \frac{d^2R}{dr^2} > 0 \; ,
\end{eqnarray} 
for the extremum to be stable. This implies that the orbit will be marginally stable if $ R^{\prime\prime} = 0$. Thus, the three equations $R = R^\prime = R^{\prime\prime} = 0$ can be numerically solved for a given $\iota_{\rm LSO}$ to yield $r$, $E$, $L_z$ and $Q$ at the LSO.

\subsection{The constants in the transition regime}
\label{sec:circ_const}
We need a model of the phase space trajectory, $[E(t),L_z(t),Q(t)]$ near the LSO in order to compute the world line of the compact object as it transitions from inspiral to plunge.  To this end, we Taylor expand  about the LSO to obtain
\begin{eqnarray}
\label{eq:E(t)}
E(t) & \simeq & E_{\rm LSO} +  (t-t_{\rm LSO}) \dot E_{\rm LSO} \;,\\
\label{eq:Lz(t)}
L_z(t) & \simeq & L_{z,\rm LSO} +  (t-t_{\rm LSO}) \dot L_{z,\rm LSO} \;,\\
\label{eq:Q(t)}
Q(t) & \simeq & Q_{\rm LSO} +  (t-t_{\rm LSO}) (\dot Q_{\rm LSO} + \dot{\delta Q}) \nonumber\\
	&  & + \delta Q\;,
\end{eqnarray}
which are natural generalizations of equations (3.4) and (3.5) of Ref.\ \cite{ot2000}. The overdot denotes differentiation with respect to $t$. We will later see that our initial condition for $t$ amounts to choosing $t_{\rm LSO}$, the instant at which the compact object crosses the LSO. This choice is consistent with the procedure in Ref.\ \cite{ot2000} --- Eq.\ (3.14) of Ref.\ \cite{ot2000} implies a choice of $t_{\rm LSO} =0$.

The constant terms in Eq.\ (\ref{eq:Q(t)}), $\delta Q$ and $\delta\dot Q$, are needed to
guarantee that the trajectory remains circular as we enter the transition. As the notation suggests, these constants are small compared to $Q_{\rm LSO}$ and $\dot Q_{\rm LSO}$. They are
discussed in more detail when we discuss initial conditions for the transition in Sec.\ \ref{sec:ics}.

The expressions (\ref{eq:E(t)}), (\ref{eq:Lz(t)}) and (\ref{eq:Q(t)}) do not include conservative effects of the self force. Pound and Poisson \cite{pp} have demonstrated that this omission will lead to observationally significant changes. Inclusion of these effects would effectively alter the potentials, Eq.\ (\ref{eq:R}) - Eq.\ (\ref{eq:Vt})  leading to slight deviations of $(E,L_z,Q)_{\rm LSO}$ and $r_{\rm LSO}$ (for a given $\iota_{\rm LSO}$) from their geodesic values. The exact impact of these effects will not be known until we know what the corrections are. We will later see that our results posses all the expected qualitative features despite this handicap. Moreover, the prescription in \cite{ot2000} and its generalization presented here can easily incorporate these effects once they are known.  

The fluxes at the LSO remain a parameter in our code. We use the code developed in \cite{hughes2000} to provide us the dimensionless fluxes, $(M/\mu)^2\dot E$, $(M/\mu^2)\dot L_z$ and $(1/\mu^3)\dot Q$ at the LSO. Equivalently, we can use the expressions in \cite{GG} (with zero eccentricity) for the dimensionless fluxes.

\subsection{Reparametrization of the $\theta$-equation}
Numerical integration of the $\theta$-equation warrants some care. The issue arises because $d\theta/dt$ vanishes at the turning points, $\theta_{\rm max}$ and $\theta_{\rm min}$, where 
\begin{equation}
0 \leq \theta_{\rm min} \leq \theta_{\rm max} \leq \pi \; .
\end{equation}
The potential problems posed by the turning points can be eliminated by reparametrizing $\theta$. Following Ref.\ \cite{hughes2000}, we use
\begin{eqnarray}
z = \cos^2\theta & = & z_-\cos^2\chi \;,
\end{eqnarray}
where
\begin{eqnarray}
\beta(z - z_+)(z-z_-) & = & \beta z^2 - z\frac{Q + L_z^2 + a^2(\mu^2-E^2)}{\mu^2} \nonumber  \\
	&  &  + \frac{Q}{\mu^2} \;,
\end{eqnarray}
and $\beta = a^2(\mu^2-E^2)/\mu^2$. The $\theta$-equation of motion now becomes
\begin{eqnarray}
\label{eq:chi}
\frac{d\chi}{dt} = \frac{\sqrt{\beta(z_+-z)}}{\gamma + a^2 E z(\chi)/\mu} \;, 
\end{eqnarray}
where
\begin{eqnarray}
\gamma = \frac{E}{\mu}\left[\frac{(r^2+a^2)^2}{\Delta} - a^2\right] - \frac{2 M r aL_z}{\Delta \mu} \;.
\end{eqnarray}
Equation (\ref{eq:chi}) can now be integrated without turning points because $\chi$ varies from $0$ to $\pi$ to $2\pi$ as $\theta$ varies from $\theta_{\rm min}$ to $\theta_{\rm max}$ and back to $\theta_{\rm min}$. 

\subsection{The prescription}
In keeping with our main objective of obtaining the world line $[r(t),\theta(t),\phi(t)]$ through the transition regime, we eliminate $\tau$ by dividing equation (\ref{geod1}) by   (\ref{geod4}) and squaring the result to obtain
\begin{eqnarray}
\left(\frac{dr}{dt}\right)^2 = \frac{R(r,\chi)}{V_t(r,\chi)^2} \equiv F\;.
\end{eqnarray}
One more time derivative gives the acceleration:
\begin{equation}
\label{eq:accn1}
\frac{d^2r}{dt^2} = \frac{1}{2}\left[\frac{\partial}{\partial r}\left(\frac{R}{V_t^2}\right) + \frac{\partial}{\partial \chi}\left(\frac{R}{V_t^2} \right)\frac{d\chi/dt}{dr/dt} \right] \;.
\end{equation}
Ideally, Eq.\ (\ref{eq:accn1}) must have other additive terms proportional to non-zero powers of $\mu$. This is analogous to Eq.\ (3.10) of \cite{ot2000}. Excluding this term amounts to ignoring the conservative self force. 

Since the transition phase is in the proximity of the LSO, we can Taylor expand $F$ about $r_{\rm LSO}$, $E_{\rm LSO}$, $L_{z,\rm LSO}$ and $Q_{\rm LSO}$ to obtain  
\begin{widetext}
\begin{eqnarray}
\label{eq:Ftexp}
F(r,L_z,E,\chi,\iota) & \simeq & \frac{1}{6}\left.\frac{\partial^3 F }{\partial r^3}\right|_{\rm LSO} (r-r_{\rm LSO})^3 + \left.\frac{\partial^2 F}{\partial r\partial L_z}\right|_{\rm LSO}(L_z-L_{z,\rm LSO})(r-r_{\rm LSO}) \nonumber \\   
	 &  &+ \left.\frac{\partial^2 F}{\partial r\partial E}\right|_{\rm LSO}(E - E_{\rm LSO})(r-r_{\rm LSO}) + \left.\frac{\partial^2 F}{\partial r\partial Q}\right|_{\rm LSO}(Q - Q_{\rm LSO})(r-r_{\rm LSO})  \;.
\end{eqnarray}
Thus, the acceleration now becomes \footnote{Note that Eq.\ (\ref{eq:Ftexp})  ignores terms of order $(\mu/M)^2$ and higher.}:
\begin{eqnarray}
\label{eq:accn}
\frac{d^2r}{dt^2} & = &  \frac{1}{2}\left[\frac{1}{2}\left.\frac{\partial^3 F}{\partial r^3}\right|_{\rm LSO} (r-r_{\rm LSO})^2 + \left.\frac{\partial^2 F}{\partial r\partial L_z}\right|_{\rm LSO}(L_z-L_{z,\rm LSO}) + \left.\frac{\partial^2 F}{\partial r\partial E}\right|_{\rm LSO}(E - E_{\rm LSO}) +\left.\frac{\partial^2 F}{\partial r\partial Q}\right|_{\rm LSO}(Q - Q_{\rm LSO}) \right.\;\nonumber \\
	&  &  \left. + \frac{\partial F} {\partial \chi}\frac{d\chi/dt}{dr/dt}\right] \;.
\end{eqnarray}
\end{widetext}
We have not expanded the second term in Eq.\ (\ref{eq:accn1}) because we do not know the value of $\chi$ at $r=r_{\rm LSO}$ a priori. Similarly, the $\phi$-equation takes the form
\begin{equation}
\label{eq:phi}
\frac{d\phi}{dt}   = \frac{V_\phi(r,\chi)}{V_t(r,\chi)}\;.
\end{equation} 
The trajectory in the transition phase can now be computed by integrating equations (\ref{eq:accn}), (\ref{eq:phi}) and (\ref{eq:chi}) from some starting point outside the LSO to  some ending point inside the LSO, for a given $\iota_{\rm LSO}$, with time varying $E$, $L_z$ and $Q$. 

\subsection{Initial conditions}
\label{sec:ics}
The angles, $\phi$ and $\chi$ can be set to zero without loss of generality. Setting $\chi = 0$ corresponds to starting the inspiral at $\theta = \theta_{\rm min}$. 

The choice of initial radius depends explicitly on $\mu$. In Ref. \cite{ot2000}, the authors define parameters, $\alpha$, $\beta$, $\kappa$, $\tau_0$ and $R_0$. These are used to scale out the perturbing mass from the equation of motion and initial conditions. Although we prefer to retain dimensions in the equations of motion, we specify initial conditions in a dimensionless form, independent of $\mu$. This will be useful be in interpreting our results and making comparisons with  Ref.\ \cite{ot2000}. Following Ref. \cite{ot2000}, we define
\begin{eqnarray}
\label{eq:X}
X & = & \left(\frac{\mu}{M}\right)^{2/5}\frac{r-r_{\rm LSO}}{R_0} \; , \\
\label{eq:R0}
R_0 & = & (\beta\kappa_0)^{2/5} \alpha^{-3/5} \; , \\
T & = & \left(\frac{\mu}{M}\right)^{1/5} \left. \frac{\tilde t-\tilde t_{\rm LSO}}{\tau_0} \frac{d\tau}{dt}\right|_{\rm LSO}  \;,
\end{eqnarray}
where
\begin{eqnarray}
\alpha  &  = &  -\frac{1}{4}\frac{\partial^3}{\partial \tilde r^3}\left[\frac{R}{\Sigma^2}\right]_{\rm LSO}   \;,\\
\beta & = &  \frac{1}{2}\left[\frac{\partial^2}{\partial \tilde L_z\partial \tilde r}\left(\frac{R}{\Sigma^2}\right) + \frac{\dot{\tilde E}}{\dot {\tilde {L_z}}}\frac{\partial^2}{\partial \tilde E\partial \tilde r}\left(\frac{R}{\Sigma^2}\right) \right.  \nonumber \\
	&  & \left. + \frac{\dot{\tilde Q}}{\dot {\tilde {L_z}}}\frac{\partial^2}{\partial\tilde Q\partial \tilde r}\left(\frac{R}{\Sigma^2}\right) \right]_{\rm LSO}  \\
\kappa(t) & = & -\frac{1}{\mu/M}\frac{d\tilde L_z}{d\tilde \tau} = - \frac{d\tilde L_z/d\tilde t}{(\mu/M)(d\tau/dt)} \;,\\
  \kappa_0 & = &  \kappa |_{\rm LSO} \;,\\
\label{eq:tau0}
\tau_0 & = & (\alpha \beta \kappa_0)^{-1/5}\;,
\end{eqnarray}
with $\tilde r = r/M$, $\tilde t = t/M$, $\tilde E = E/\mu$, $\tilde L_z = L_z/(\mu M)$ and $\tilde Q = Q/(\mu M)^2$.

These definitions reduce to those presented in Ref.\ \cite{ot2000} when $\iota=0$. It is useful to observe that $\kappa$ does not scale with $\mu$. We evaluate $d\tau/dt$, $\alpha$, $\beta$ and $\kappa_0$ at $\theta = \pi/2 - \iota_{\rm LSO}$  because we do not know $\theta_{\rm LSO}$ a priori.  Notice that $X$ and $T$ are dimensionless. 

The smoothness of the transition implies that there is no fixed instant at which the transition starts or ends. Motivated by the choices in Ref.\ \cite{ot2000}, we set  $T \simeq -1$  at $t=0$  and stop the numerical integrator when $X \leq X_e = -5$.   

In summary, our initial conditions are $T = -1$, $\phi=0$ and $\chi = 0$ at $t=0$ \footnote{It is important to keep $|T|$ small enough that our Taylor expansion about the LSO remains a valid approximation.}. Setting $T=-1$ at $t=0$ allows us to calculate $t_{\rm LSO}$ and hence $E(0)$ and $L_z(0)$ from equations (\ref{eq:E(t)}) and (\ref{eq:Lz(t)}). We then solve $R(E,L_z,Q,r)=0$ and $dR/dr =0$ to obtain $r(0)$ and $Q(0)$. This is analogous to Sec.\ IIIC of Ref.\ \cite{ot2000} where they enforce $X=\sqrt{-T}$ to determine $X$ at $t=0$. 

The trajectory is adiabatic before the start of the transition. At $t=0$, we must impose the condition \cite{circ1,circ2,circ3} that circular orbits remain circular even under adiabatic radiation reaction. Thus, requiring that $\dot R = dR/dt = 0$ and $\dot R^\prime = d^2R/drdt = 0$ leads to expressions (3.5) and (3.6) of \cite{hughes2000} for $\dot r(0)$ and $\dot Q(0)$ respectively.  

We can now substitute $Q(0)$ and $\dot Q(0)$ in Eq.\ (\ref{eq:Q(t)}) to obtain two independent equations,
\begin{eqnarray}
Q(0) & = & Q_{\rm LSO} -t_{\rm LSO}(\dot Q_{\rm LSO} + \delta \dot Q)  + \delta Q \; \mbox{and}\\
\dot Q(0)  & = & \dot Q_{\rm LSO} + \delta \dot Q \;, 
\end{eqnarray}
which can be used to evaluate $\delta Q$ and $\dot{\delta Q}$.

\subsection{Code algorithm and numerical results}
The previous sections developed the steps required to calculate the compact body's trajectory as it transitions from inspiral to plunge. We now summarize the algorithm that was actually used to implement this prescription:
\newline
(1) Take $\iota_{\rm LSO}$ as input.\newline
(2) Compute $E$ and $L_z$  at the LSO. \newline
(3) Obtain $\dot E$ and $\dot L_z$ at the LSO from the code developed in \cite{hughes2000}. We may also use the expressions in \cite{GG} (which reduce to the results in
\cite{hughes2000} for circular orbits), which will be particularly useful when we
generalize to eccentric orbits.\newline
(4) Choose initial conditions $T \simeq -1$, $\phi = 0$ and $\chi = 0$ at $t=0$. \newline
(5) Calculate $E(0)$ and $L_z(0)$ from equations (\ref{eq:E(t)}) and (\ref{eq:Lz(t)}). \newline
(6) Solve for $r(0)$ and $Q(0)$ by imposing $R=0$ and $dR/dr=0$ at $t=0$.\newline
(7) Compute $\dot r(0)$ and $\dot Q(0)$ from equations (3.5) and (3.6) of \cite{hughes2000}. \newline
(8) Substitute $Q(0)$ and $\dot Q(0)$ in Eq.\ (\ref{eq:Q(t)}) to evaluate $\delta Q$ and $\delta \dot Q$. \newline
(9) Use a Runge-Kutta integrator on (\ref{eq:accn}), (\ref{eq:phi}) and (\ref{eq:chi}) to compute the coordinates at the next step. A time step of $\delta t \simeq 0.05 M$ works well. \newline
(10) Update the ``constants'', $E_{i+1}  = E_i +\dot E \delta t$,  $L_{z,i+1} = L_{z,i} +\dot L_z \delta t$ and $Q_{i+1} = Q_i + (\dot Q + \dot{\delta Q}) \delta t$. The subscript $i$ denotes a discrete time instant.\newline
(11) Repeat steps (9)-(11) until $X(t) \simeq -5$.

\begin{figure}[htb]
\begin{center}
\includegraphics[height = 70mm]{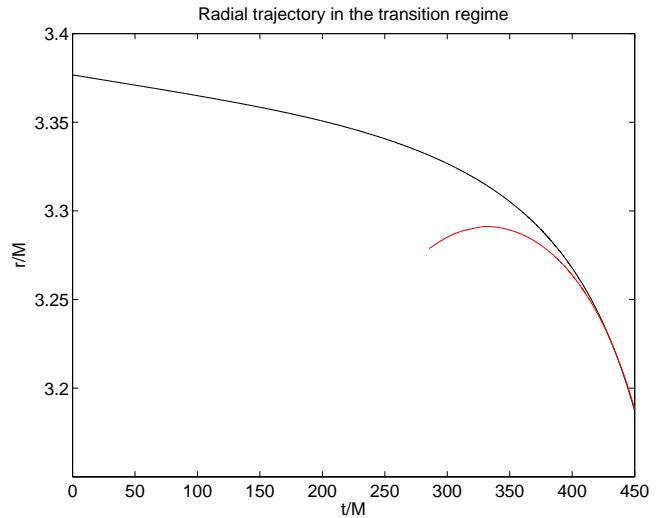}
\caption{\label{fig:circ_r}Radial trajectory during the transition (black line) from inspiral to plunge for a compact object of mass $\mu = 10^{-5}M$ in a nearly circular orbit around a black hole with spin $a=0.8M$. The compact object crosses the LSO at time $t_{\rm LSO} = 137.5M$. The inclination of the orbit at $t_{\rm LSO}$ is $\iota_{\rm LSO} = 37^\circ$. The red line is a plunging geodesic matched to the end of the transition.}
\end{center}
\end{figure}

\begin{figure}[htb]
\begin{center}
\includegraphics[height = 70mm]{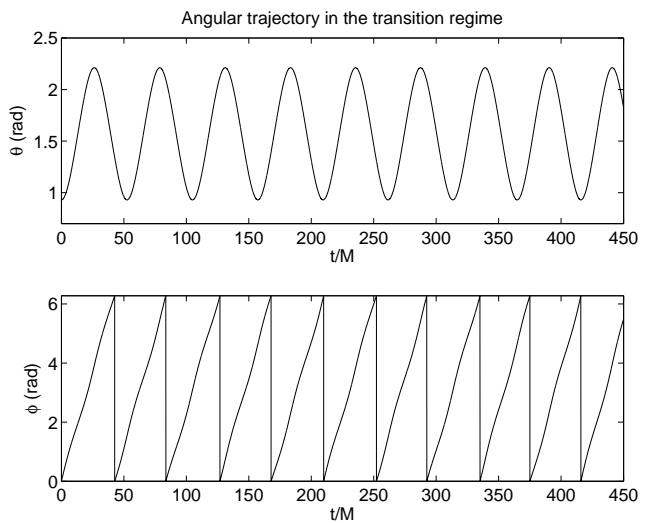}
\caption{\label{fig:circ_ang}Angular motion during the transition for a compact object around a spinning black hole with identical parameters as in Fig.\ \ref{fig:circ_r}.}
\end{center}
\end{figure}

The primary objective of this calculation is to compute the world line of the compact object during the transition. Figures \ref{fig:circ_r} and \ref{fig:circ_ang} illustrate $r$, $\theta$ and $\phi$ motions of the compact object for a typical set of parameters. We also show a plunging geodesic matched to the end of the transition.

Table\ \ref{tbl:circ} shows the parameters and transit times for a range of inclination angles. In general, we find that the transit time increases with inclination. However, the dimensionless transit time $\Delta T$ remains approximately constant,
\begin{equation}
\Delta T \simeq 3.3 - 3.4 \; ,
\end{equation}
when $X_e = -5$ for all values of $a$ and $\iota$. Again, this is a consistent generalization of the result in Ref.\ \cite{ot2000} where they find $\Delta T \simeq 3.3$ for all circular, equatorial orbits. 
 
\begin{table*}[htb]
\caption{\label{tbl:circ} Fluxes and transit times for different inclinations. We set $a=0.5M$, $\mu = 10^{-6}M$ , $M=1$, $T(0)=-1$ and $X_e = -5$.}  
\begin{tabular}{c|c|c|c|c|c|c|c|c|c|c|c}  
\hline  
\hline  
$\iota_{\rm LSO} ^\circ$ & $r_{\rm LSO}/M$ & $(M/\mu)^2\dot E_{\rm LSO}$ & $(M/\mu^2)\dot L_{z,\rm LSO}$ & $(1/\mu^3)\dot Q_{z,\rm LSO}$ & $\alpha$ & $\beta$ & $R_0$ & $\kappa_0$ & $\tau_0$ & $t/M$ & $\Delta T$ \\  
\hline
$10^{-3}$ & $4.23$ & $-0.00457$ & $-0.0422$ & $-0.000572$ & $0.00311$ & $0.0327$ & $0.0699$ & $0.603$ & $2.80$ & $944.9$ & $3.36$ \\
$10$ & $4.26$ & $-0.00446$ & $-0.0409$ & $-0.00684$ & $0.00304$ & $0.0327$ & $0.0677$ & $0.604$ & $2.81$ & $952.4$ & $3.36$\\
$20$ & $4.32$ & $-0.00415$ & $-0.0375$ & $-0.0241$ & $0.00284$ & $0.0329$ & $0.0615$ & $0.610$ & $2.82$ & $974.9$ & $3.36$ \\
$30$ & $4.43$ & $-0.00368$ & $-0.0323$ & $-0.0481$ & $0.00254$ & $0.0333$ & $0.0523$ & $0.618$ & $2.84$ & $1012.6$ & $3.36$\\
$40$ & $4.59$ & $-0.00314$ & $-0.0262$ & $-0.0733$ & $0.00219$ & $0.0342$ & $ 0.0416$ & $0.630$ & $2.86$ & $1065.9$ & $3.35$ \\
$50$ & $4.78$ & $-0.002594$ & $-0.0198$ & $-0.0946$ & $0.00184$ & $0.0363$ & $0.0309$ & $0.643$ & $2.88$ & $1134.4$ & $3.35$ \\
$60$ & $5.01$ & $-0.00208$ & $-0.0139$ & $-0.108$ & $0.00152$ & $0.0403$ & $0.0211$ & $0.657$ & $2.90$ & $1217.9$ & $3.35$\\
\hline    
\hline  
\end{tabular}  
\end{table*} 

\begin{table}[htb]
\caption{\label{tbl:mu} Variation of transit time with perturbing mass, $\mu/M$. We set $a=0.9M$, $\iota_{\rm LSO} = 0.001^\circ$, $M=1$, $T_s=-1$ and $X_e = -5$. Note that $r_{\rm LSO} = 2.32M $.}  
\begin{tabular}{c|c|c}
\hline  
\hline
$\mu/M$ & $t/M$ & $\Delta T$\\
\hline
$10^{-3}$ & $118.9$ & $3.449$ \\
$10^{-4}$ & $185.6$ & $3.397$ \\
$10^{-5}$ & $292.2$ & $3.375$ \\
$10^{-6}$ & $461.9$ & $3.367$ \\
$10^{-7}$ & $731.3$ & $3.363$ \\
$10^{-8}$ & $1158.6$ & $3.362$ \\
\hline
\hline
\end{tabular}  
\end{table} 

\subsection{Comparison with Ref.\ \cite{ot2000}}
The results in Ref.\ \cite{ot2000} provide an important sanity check for the case of circular, equatorial orbits. However, we have to account for the minor differences between the two approaches. Ref.\ \cite{ot2000} makes the approximations
\begin{eqnarray}
\frac{d\phi}{dt} & \simeq & \left. \frac{d\phi}{dt}\right|_{\rm ISCO} \; \;  \mbox{and} \\
\frac{d\tau}{dt} & \simeq & \left. \frac{d\tau}{dt}\right|_{\rm ISCO} \;  ,
\end{eqnarray}
which lead to 
\begin{eqnarray}
 \kappa & = & - \frac{d\tilde L_z/d\tilde t}{(\mu/M)(d\tau/dt)} \;,\nonumber\\
 	& \simeq & - \frac{d\tilde L_z/d\tilde t|_{\rm ISCO}}{(\mu/M)(d\tau/dt)_{\rm ISCO}} \;,
\end{eqnarray}
which is a dimensionless constant. In our prescription, $d\tau/dt$ varies with time. This time dependence has to be enforced because $d\tau/dt$ is a function of $\theta$, whose value at the LSO is not known a priori. The circular, equatorial case in Ref.\ \cite{ot2000} does not suffer from  this pathology because $\theta = \pi/2$ at all times. Thus, we treat $\kappa$ as a slowly varying function of time. Table\ \ref{tbl:mu} shows the transit times for a nearly equatorial orbit ($\iota_{\rm LSO} = 0.001$) and a range of mass ratios. As the mass ratio becomes smaller, the variation in $\kappa$ decreases, and the dimensionless transit time converges to the limit where $\kappa$ is constant.  

Our initial conditions differ slightly from those used in Ref.\ \cite{ot2000}. Effectively, they use the Taylor expansion of $R(r)$ to solve $dR/dr = 0$ and $d^2R/(drdt) = 0$ for $r(0)$ and $\dot r(0)$ respectively. In contrast, we solve the equations exactly. This leads to  differences of less than $1\%$.

\section{Eccentric orbits}
\label{sec:ecc}
The methods developed thus far only discussed circular orbits. We now extend this technique to include non-zero eccentricity. In the absence of radiation reaction, the geodesic equations admit bound eccentric orbits. These orbits are conventionally parametrized by the semi-latus rectum, $p$, and the eccentricity, $e$. The radial coordinate can now be expressed as
\begin{equation}
r(t) = \frac{p}{1+e\cos\psi(t)}\;.
\end{equation}
The angle $\psi(t)$ is analogous to the eccentric anomaly and can be solved for numerically. The geodesic has turning points at $\psi = 0$,$\pi$. Deep in the adiabatic inspiral, the compact object's trajectory is well approximated by a sequence of orbits with slowly varying $p(t)$ and $e(t)$. 

Geodesics beyond the LSO do not have turning points (where $dr/dt=0$). This changes the situation considerably because the parameters, $p$ and $e$ are not well-defined anymore. Thus, the trajectory ceases to have turning points somewhere during the transition from inspiral to plunge. We will later show that this feature is naturally buried in our model of the transition.

\subsection{The last stable orbit}
\label{sec:lso}
As with circular orbits, the last stable bound geodesic is an important reference in our procedure. The inner and outer turning points ($r_{\rm min}$ and $r_{\rm max}$) of the LSO are related to $e_{\rm LSO}$ and $p_{\rm LSO}$ through
\begin{eqnarray}
r_{\rm min} & = & \frac{p_{\rm LSO}}{1+e_{\rm LSO}} \; \mbox{and}\\
r_{\rm max} & = & \frac{p_{\rm LSO}}{1-e_{\rm LSO}} \;.
\end{eqnarray}
Our goal is to determine $p_{\rm LSO}$ and the constants $(E,L_z,Q)$ at the LSO for a given $\iota_{\rm LSO}$ and $e_{\rm LSO}$. This can be achieved by requiring that 
\begin{eqnarray}
\label{eq:lso1}
\frac{dR}{dr}  & = & 0 \mbox{ at } r = r_{\rm min} \; , \\
\label{eq:lso2}
R  & = & 0 \mbox{ at } r = r_{\rm min} \mbox{ and } r = r_{\rm max} \;.
\end{eqnarray}
Recall that the function $R$ is given by Eq.\ (\ref{eq:R}) and $\iota$ is defined by Eq.\ (\ref{eq:Q}). We require Eq.\ (\ref{eq:lso1}) to be satisfied because the inner most turning point corresponds to a local maximum of $(-R)$. Equation (\ref{eq:lso2}) enforces the compact object's velocity to vanish at the turning points.  Equations (\ref{eq:lso1}), (\ref{eq:lso2}) and (\ref{eq:Q}) can be solved numerically for $p$ and $(E,L_z,Q)$ at the LSO. Appendix\ \ref{app:lso} describes the details of this numerical procedure.  

\subsection{The constants during the transition}
As with the circular case, our initial conditions are such that we effectively {\em{choose}} the LSO crossing to occur at $t=t_{\rm LSO}$. This allows us to expand the constants about the LSO to obtain
\begin{eqnarray}
\label{eq:E(t)ecc}
E(t) & \simeq & E_{\rm LSO} +  (t-t_{\rm LSO}) \dot E_{\rm LSO} \;,\\
\label{eq:Lz(t)ecc}
L_z(t) & \simeq & L_{z,\rm LSO} +  (t-t_{\rm LSO}) \dot L_{z,\rm LSO} \;,\\
\label{eq:Q(t)ecc}
Q(t) & \simeq & Q_{\rm LSO} +  (t-t_{\rm LSO}) \dot Q_{\rm LSO} .
\end{eqnarray}
Notice that we no longer need the corrections, $\delta Q$ and $\dot{\delta Q}$ because there are no additional symmetries to constrain $Q(0)$ and $\dot Q(0)$ ; $E(t)$, $L_z(t)$ and $Q(t)$ are independent. 

As discussed in Sec.\ \ref{sec:circ_const}, equations (\ref{eq:E(t)ecc}), (\ref{eq:Lz(t)ecc}) and (\ref{eq:Q(t)ecc}) do not include conservative effects of the self force. Just as the circular case, this will lead to a slight shift of $(E,L_z,Q)_{\rm LSO}$ and $p_{\rm LSO}$ (for a given $e_{\rm LSO}$ and $\iota_{\rm LSO}$) with respect to their geodesic values. Again, our motivation to stick with this approximation stems from the facts that: (a) These effects can be incorporated into our prescription once they are known, and (b) Our results show the generally expected behavior, at least qualitatively.
 
Numerical methods to calculate the change in the Carter constant due to
gravitational-wave backreaction have recently become available  \cite{qdot1,qdot2}. Work is in
progress implementing that result in the code we use to compute the rate of change of
orbital constants \cite{qdot3}.  For now, we use the approximate expressions for $\dot Q$
described in \cite{GG}; it will be a simple matter to update our code when more accurate $\dot Q$ results are available.

\subsection{The prescription for eccentric orbits}
\label{sec:ecc_pres}
Our next task is to  derive equations of motion to map the phase space trajectory to an actual world line. The angular equations, Eq.\ (\ref{eq:chi}) and Eq.\ (\ref{eq:phi}), remain unaffected. Our strategy for the radial equation is to expand the geodesic equation about $(E_{\rm LSO},L_{z,\rm LSO},Q_{\rm LSO})$. This leaves us with
\begin{widetext}
\begin{eqnarray}
\label{eq:accnecc}
\frac{d^2r}{dt^2} & = & \frac{1}{2}\left[\frac{\partial F}{\partial r} + \frac{\partial F}{\partial \chi} \frac{d\chi/dt}{dr/dt}\right] \;,\\
\label{eq:accn3}
\frac{\partial F}{\partial r} & \simeq & \left[ \left.\frac{\partial^2 F}{\partial r \partial E}\right|_{\rm LSO} (E-E_{\rm LSO}) + \left.\frac{\partial^2 F}{\partial r \partial L_z}\right|_{\rm LSO}(L_z - L_{z,LSO}) + \left.\frac{\partial^2 F}{\partial r \partial Q}\right|_{\rm LSO} (Q-Q_{\rm LSO}) \right. \nonumber \\ 
	&  &\left. + \frac{\partial F}{\partial r}(r,\chi;E_{\rm LSO},L_{z,\rm LSO},Q_{\rm LSO}) \right] \;.
\end{eqnarray}
\end{widetext}
Note that we only expand about the constants, not the $r$-coordinate, because there is no unique $r$ at the LSO. In the absence of the first three terms in Eq.\ (\ref{eq:accn3}), the equation of motion is simply a geodesic at the LSO. This is consistent with our intuitive notion of ``expanding about the LSO''. The existence of turning points presents a complication while integrating Eq.\ (\ref{eq:accn3}) numerically. We present a method to tackle this in Appendix\ \ref{app:tp}.

\subsection{Initial conditions}
We need initial conditions for $r$ and $dr/dt$ before we start the numerical integrator. Motivated by the initial conditions for circular orbits, we set $T\simeq -1$ at $t=0$. This amounts to choosing $t_{\rm LSO}$. We can now determine $[E(0),L_z(0),Q(0)]$, which can be mapped to $(p, e, \iota)$ at $t=0$. This mapping is allowed because the trajectory is adiabatic before $t=0$. The coordinates at any point on the geodesic defined by $[E(0),L_z(0),Q(0)]$ can serve as our initial conditions. For simplicity, we choose
\begin{eqnarray}
r(0) & = & \frac{p}{1+e} \;, \\
\frac{dr}{dt}(0) & = & 0 \;, \\
\phi(0) & = & 0 \;, \\
\chi(0) & = & 0 \;.
\end{eqnarray} 
The equations of motion can now be easily integrated across the LSO.

\subsection{Code implementation and numerical results}
Taking eccentricity into account changes our algorithm slightly. We summarize the code's algorithm as follows:
\newline
(1) Take $\iota_{\rm LSO}$ and $e_{\rm LSO}$ as input.\newline
(2) Compute $E$, $L_z$ and $Q$ at the LSO. \newline
(3) Obtain $\dot E$, $\dot L_z$ and $\dot Q$ at the LSO from the expressions in Ref.\ \cite{GG}. \newline
(4) Choose initial conditions $T \simeq -1$, $\phi = 0$ and $\chi = 0$ at $t=0$. \newline
(5) Calculate $E(0)$, $L_z(0)$ and $Q(0)$ from equations (\ref{eq:E(t)ecc}), (\ref{eq:Lz(t)ecc}) and (\ref{eq:Q(t)ecc}). \newline
(6) Map  $\left[E(0),L_z(0), Q(0)\right]$ to $(p,e,\iota)$. \newline
(7) Set $r = p/(1+e)$ and $dr/dt = 0$ at $t = 0$. \newline
(8) Use a Runge-Kutta integrator on (\ref{eq:accnecc}), (\ref{eq:phi}) and (\ref{eq:chi}) to compute the coordinates at the next step. A time step of $\delta t \simeq 0.05 M$ works well. \newline
(10) Update the ``constants'', $E_{i+1}  = E_i +\dot E_{\rm LSO} \delta t$,  $L_{z,i+1} = L_{z,i} +\dot L_{z, \rm LSO} \delta t$ and $Q_{i+1} = Q_i + \dot Q_{\rm LSO} \delta t$. The subscript $i$ refers to a discrete time instant.\newline
(11) Repeat steps (9)-(11) until $X \simeq -5$.

Recall that the local minimum of the potential $R$ is less than zero for bound orbits and is greater than zero for a plunging geodesic. The minimum is exactly zero at the LSO. These conditions can be used as sanity checks while performing the numerical integration. 

\begin{table*}[htb]
\caption{\label{tbl:gen} Fluxes and transit times for different eccentricities. We set $a=0.8M$, $\mu = 10^{-5}$, $\iota_{\rm LSO} = 45^\circ$, $M=1$, $T_s=-1$ and $X_e = -5$.}  
\begin{tabular}{c|c|c|c|c|c|c|c|c|c|c|c}  
\hline  
\hline  
$e_{\rm LSO}$ & $p_{\rm LSO}/M$ & $(M/\mu)^2\dot E_{\rm LSO}$ & $(M/\mu^2)\dot L_{z,\rm LSO}$ & $(1/\mu^3)\dot Q_{z,\rm LSO}$ & $\alpha$ & $\beta$ & $R_0$ & $\kappa_0$ & $\tau_0$ & $t/M$ & $\Delta T$ \\  
\hline
$10^{-4}$ & $3.58$ & $-0.00974$ & $-0.0619$ & $-0.153$ & $0.00517$ & $0.0530$ & $3.04$ & $0.113$ & $7.98$ & $486.3$ & $3.34$ \\
$0.1$ & $3.70$ & $-0.00857$ & $-0.0545$ & $-0.136$ & $0.00351$ & $0.0506$ & $3.54$ & $0.0969$ & $8.97$ & $448.2$ & $2.81$ \\
$0.2$ & $3.84$ & $-0.00795$ & $-0.0479$ & $-0.120$ &  $0.00220$ & $0.0484$ & $4.33$ & $0.0832$ & $10.2$ & $373.5$ & $2.10$  \\
$0.3$ & $3.96$ & $-0.00751$ & $-0.0419$ & $-0.105$ & $0.00117$ & $0.0463$ & $5.83$ & $0.0714$ & $12.1$ &  $341.1$  & $1.66$\\  
$0.4$ & $4.09$ & $-0.00693$ & $-0.0361$ & $-0.0900$ & $0.000365$ & $0.0442$ & $10.8$ & $0.0604$ & $15.9$ & $332.7$ & $1.25$ \\
$0.5$ & $4.22$ & $-0.00607$ & $-0.0300$ & $-0.0745$ & $-0.000280$ & $0.0420$ & $11.5$ & $0.0496$ & $17.7$ &  $331.6$ & $1.14$ \\
$0.6$ & $4.35$ & $-0.00450$ & $-0.0236$ & $-0.0582$ & $-0.000801$ & $0.0401$ & $5.41$ & $0.0385$ & $15.2$ & $338.8$ & $1.37$  \\
$0.7$ & $4.49$ & $-0.00351$ & $-0.0168$ & $-0.0413$ & $-0.00123$ & $0.0381$ & $3.57$ & $0.0272$ & $15.1$ & $381.9$ & $1.56$ \\
$0.8$ & $4.62$ & $-0.00206$ & $-0.0100$ & $-0.0245$ & $-0.00159$ & $0.0362$ & $2.44$ & $0.0162$ & $16.1$ & $507.2$ & $1.95$ \\
\hline    
\hline  
\end{tabular}  
\end{table*} 

\begin{figure}[htb]
\begin{center}
\includegraphics[height = 70mm]{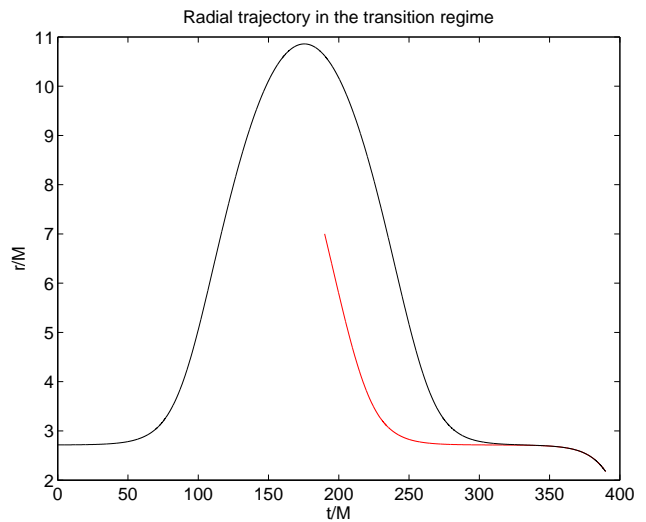}
\caption{\label{fig:gen_r}
Radial trajectory during the transition (black line) from inspiral to plunge for a compact object of mass $\mu = 10^{-6}M$ in an eccentric orbit around a black hole with spin $a=0.8M$. The compact object crosses the LSO at time $t_{\rm LSO} = 196.7M$. The inclination and eccentricity of the orbit at $t_{\rm LSO}$ are $\iota_{\rm LSO} = 45^\circ$ and $e_{\rm LSO} = 0.6$ respectively. The red line is an unstable geodesic matched to the end of the transition.}
\end{center}
\end{figure}

\begin{figure}[htb]
\begin{center}
\includegraphics[height = 70mm]{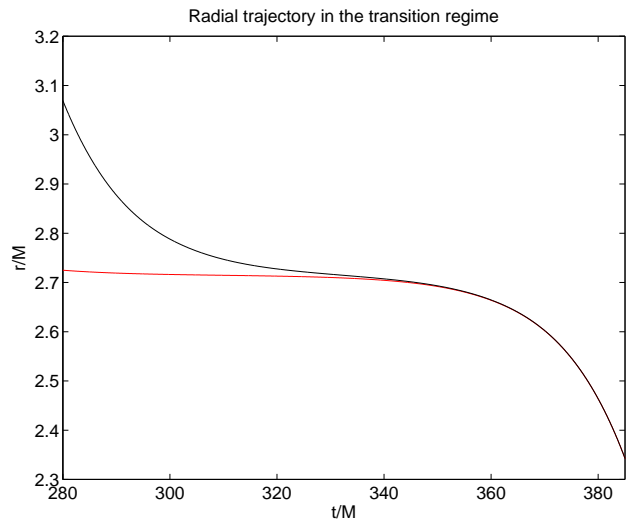}
\caption{\label{fig:gen_rzoom}
Same as Fig.\ \ref{fig:gen_r}, but zooming in on the final ``whirls''.}
\end{center}
\end{figure}

\begin{figure}[htb]
\begin{center}
\includegraphics[height = 70mm]{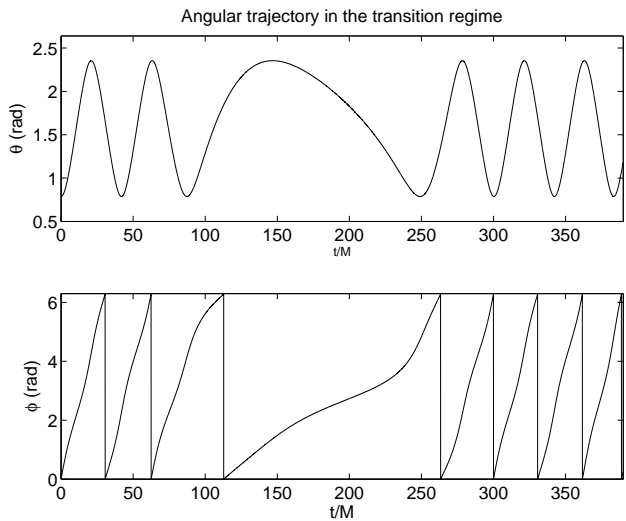}
\caption{\label{fig:gen_ang}Angular trajectory during the transition for the same set of parameters as in Fig.\ \ref{fig:gen_r}.}
\end{center}
\end{figure}

Figures \ref{fig:gen_r}, \ref{fig:gen_rzoom}  and \ref{fig:gen_ang} show a typical trajectory during the transition from inspiral to plunge. The compact object starts at the minimum of the last bound geodesic before the plunge. The radial coordinate increases until it reaches a maximum where $R = dr/dt = 0$. Subsequently, it turns around and heads toward the minimum. After executing a number of ``whirls" near the minimum, the trajectory becomes unstable, and thus plunges into the central black hole. The whirls are evident from the angular trajectory plotted in Fig.\ \ref{fig:gen_ang}. We also show a plunging geodesic matched to the end of the transition. Notice that the plunge spends quite a bit of time at $r \sim 2.8M$ --- much more time than the transition trajectory.  This is because the radiation emission built into the transition trajectory's construction pushes it off this marginally stable orbit rather quickly.

Table\ \ref{tbl:gen} shows the various parameters and transit times for a range of eccentricities. Note that the parameters $\alpha$, $\beta$, $R_0$, $\kappa_0$ and $\tau_0$ (which are defined in Sec.\ \ref{sec:ics}) are evaluated at $p_{\rm LSO}$. In general, we find that the transit time is proportional to $\alpha$. This is not surprising because $\alpha$ is the first term in the Taylor expansion of the potential, $R$. We also observe some degree of correlation between the transit time and $\tau_0$, the parameter used to define the dimensionless time. 

\subsection{Comparison with Ref.\ \cite{eq_tr}}
As mentioned in the introduction, there are differences between our generalized prescription and the method developed in Ref.\ \cite{eq_tr}, which only models the transition when the compact object is in an eccentric, equatorial orbit. First, we  set our initial conditions at the {\em start} of the LSO, whereas Ref.\ \cite{eq_tr} sets the initial conditions at the {\em end} of the LSO. This educated choice allows Ref.\ \cite{eq_tr} to derive an analytic form for the trajectory. Second, we differ in the choice of final conditions. \footnote{See Sec.\ IID{\it 3}  and Ref.\ [20] of Ref.\ \cite{eq_tr} for a description their choice of final conditions.}

\begin{figure}[htb]
\begin{center}
\includegraphics[height = 70mm]{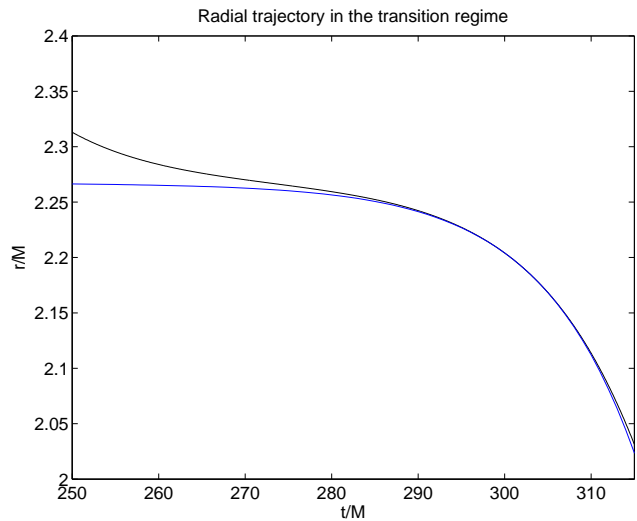}
\caption{\label{fig:eqtr_comp}Comparison of our trajectory with approximate analytic results from Ref.\ \cite{eq_tr}. The compact object is in an eccentric, equatorial trajectory with parameters $e_{\rm LSO} = 0.6$ and $\mu=10^{-6}M$. Its mass is $\mu = 10^{-6}M$ and is around a black hole with spin $a=0.8M$. The black line shows our trajectory; the blue line is obtained from Ref.\ \cite{eq_tr}. The observed deviation is because the approximation in Ref.\ \cite{eq_tr} is somewhat more restrictive than ours.}
\end{center}
\end{figure}

In attempting to make comparisons with Ref.\ \cite{eq_tr}, we found a number of typographical errors. Thus, we extract the essence of the calculation in Ref.\ \cite{eq_tr} and present it in a form that (hopefully) makes the errors obvious. We start by expressing the radial geodesic equation as
\begin{eqnarray}
\label{eq:V}
\left(\frac{dr}{d\tau}\right)^2 + V(r) & = & 0 \; ,
\end{eqnarray} 
where
\begin{eqnarray}
V(r) & = & -\frac{R}{\Sigma^2}\;.
\end{eqnarray}
The orbit is unstable if the local maximum of $V(r)$ is negative. Define
\begin{eqnarray}
I = -\mbox{Max}\{V(r)\} = -V(r_{\rm max}) \;. 
\end{eqnarray}
Note that this implies $V^{\prime}(r_{\rm max}) = 0$ and $V^{\prime\prime}(r_{\rm max}) < 0$.
We Taylor expand Eq.\ (\ref{eq:V}) about the maximum of $V(r)$ corresponding to some $(E,L_z,Q)$ just beyond the LSO to get
\begin{eqnarray}
\left(\frac{d(\delta r)}{d\tau}\right)^2 + V(r_{\rm max}) + \delta r V^\prime(r_{\rm max}) & & \nonumber \\
+ \frac{1}{2}\delta r^2 V^{\prime\prime}(r_{\rm max}) & = & 0   \; , \\
\Rightarrow \gamma^2 \left(\frac{d(\delta r)}{dt}\right)^2 + \frac{1}{2}\delta r^2 V^{\prime\prime}(r_{\rm max}) = I \; \nonumber \\
\label{eq:eq_tr}
\Rightarrow \gamma^2 \left(\frac{d(\delta r)}{dt}\right)^2 - \frac{\delta r^2}{\tau_s^2} = I \;
\end{eqnarray}
where
\begin{eqnarray}
\delta r &  = & r(t) - r_{\rm max}\; ,\\
\gamma & = &  \left.\frac{dt}{d\tau}\right|_{r_{\rm max}} = \left.\frac{V_t}{\Sigma}\right|_{r_{\rm max}} \;,\\
\tau_s^2  & = & 2/|V^{\prime\prime}(r_{\rm max})| \;. 
\end{eqnarray}  
The solution of Eq.\ (\ref{eq:eq_tr}) in the regime of interest is
\begin{eqnarray}
\label{eq:req_tr}
r(t) & = & r_{\rm max} - \sqrt{I}\tau_s\sinh\left(\frac{t-t_c}{\gamma\tau_s}\right) \;,
\end{eqnarray}
where $t_c$ is an integration constant. We can compare our numerical solution with Eq.\ (\ref{eq:req_tr}) by letting the two trajectories intersect at some arbitrary instant. This freedom is equivalent to choosing initial conditions. For example, Fig.\ \ref{fig:eqtr_comp} shows the two  trajectories near $r_{\rm max}$ for which $t_c$ is chosen such that they intersect at $t=300M$. The compact object has mass $\mu=10^{-6}M$ and is in an eccentric orbit with $e_{\rm LSO} = 0.6$ around a black hole with spin $a=0.8M$. Notice that Eq.\ (\ref{eq:req_tr}) is valid only in the {\em immediate} vicinity of $r_{\rm max}$ because $(dt/d\tau)$ and $(E,L_z,Q)$ are assumed constant. Inclusion of the time-dependence of $(dt/d\tau)$ is crucial because it leads to time varying $\gamma$, which alters the natural timescale in Eq.\ (\ref{eq:req_tr}). This explains the observed deviation at large values of $|\delta r|$.

\section{Summary and Future work}
\label{sec:summ}
The primary focus of this paper is to provide an approximate model for the trajectory of a compact object as it transitions from an adiabatic inspiral to a geodesic plunge. We have presented a generalization of the procedure in Ref.\ \cite{ot2000}, where circular, equatorial orbits are treated. We derive approximate equations of motion [Eq.\ (\ref{eq:accn}) and Eq.\ (\ref{eq:accn3})] by Taylor expanding the geodesic equations about the LSO and subjecting them to evolving $E$, $L_z$ and $Q$. We can now readily integrate these equations numerically.
Figures \ref{fig:circ_r} and \ref{fig:circ_ang} show the radial and angular trajectories for a typical inclined, circular orbit.  We also plot the plunging geodesic that it transitions to. Figures \ref{fig:gen_r} and \ref{fig:gen_ang} are analogous plots for an eccentric orbit.  Our numerical experiments suggest that the transit time is correlated with $\alpha$, the coefficient of the first term in the Taylor expansion of the radial potential.  

The code developed in \cite{skh07} and \cite{skhd08} solves the Teukolsky equation  in the time-domain and thus computes gravitational waveforms for almost any given trajectory of the compact object. We intend to generate waveforms by feeding the world lines calculated using this prescription to the time-domain Teukolsky equation-solver. The resulting waveforms will be useful for LISA data analysis routines. It is also possible to use these waveforms to estimate recoil velocities from mergers of compact objects with black holes.  
 
\acknowledgments
The author is very grateful to Scott\ A.\ Hughes for invaluable guidance throughout the development of this paper. The author also thanks Gaurav Khanna for helpful discussions. The author also thanks Adrian Liu for spotting an embarassing error in an earlier version of the paper. This work was supported by NASA Grant No.\ NNG05G105G.

\appendix
\section{The LSO for eccentric orbits}
\label{app:lso}
The following set of equations need to be solved in order to compute $p$ and $(E,L_z,Q)$ at the LSO for a given inclination ($\iota$) and eccentricity, ($e$):
\begin{eqnarray}
\label{eq:R1}
R(r,E,L_z) & = & 0 \;,\\
\label{eq:R2}
R\left(r\frac{1+e}{1-e},E,L_z\right) & = & 0 \; \mbox{and} \\
\frac{dR}{dr}(r,E,L_z) & = & 0 \;.
\end{eqnarray}
Recall that $R$ is given by Eq.\ (\ref{eq:R}). The carter constant, $Q$ can be eliminated using Eq.\ (\ref{eq:Q}). Applying an iterative technique to solve the above equations directly can lead to problems because the terms that do not contain $r$ are identical in equations (\ref{eq:R1}) and (\ref{eq:R2}). We can skirt around this problem by solving the equivalent set of equations,
\begin{eqnarray}
R_1(r,E,L_z) & = & R(r,E,L_z)  =  0 \;,\\
R_2(r,E,L_z) & = & R\left(r\frac{1+e}{1-e},E,L_z\right) - R_1(r,E,L_z) \nonumber \\
    & = & 0 \; \mbox{and} \\
R_3(r,E,L_z) & = & \frac{dR}{dr}(r,E,L_z)  =  0 \;,
\end{eqnarray} 
using the standard Newton-Raphson method described in \cite{NR}. This iterative procedure takes an initial guess for the solution as input. We use
\begin{eqnarray} 
r_0 & = &  6.1(1-a/2) \; ,\\
L_{z,0} & = &  r_0 v\cos\iota \frac{1 - 2qv^3 + q^2 v^4}{\sqrt{1 - 3v^2 + 2qv^3}} \; \mbox{and} \\ 
E_0  & = & \frac{1 - 2v^2 + qv^3}{\sqrt{1 - 3v^2 + 2qv^3}} \; .
\end{eqnarray}
where $q = a/M$, $r = \sqrt{M/r}$ and $S_0 = (r_0,L_{z,0},E_0)^T$ is our initial guess for $S = (r,L_z,E)^T$. Let $S_i$ denote the solution at any given iteration. The algorithm consists of incrementing $S_i$ as follows:
\begin{eqnarray}
S_{i+1} = S_{i} + \lambda \times \delta S_{i} \;, \\
\end{eqnarray}
where 
\begin{eqnarray}
\delta S_{i} & = & J^{-1}_i B_i \; ,\\
J_i & = & \left(\begin{array}{ccc}
	 	\partial R_1/\partial r & \partial R_1/\partial E  & \partial R_1/\partial L_z \\
	 	\partial R_2/\partial r &  \partial R_2/\partial E & \partial R_2/\partial L_z \\
		 \partial R_3/\partial r &  \partial R_3/\partial E & \partial R_3/\partial L_z
	 	\end{array}\right)_i \; ,\\
B_i & = & (-R_{1,i},-R_{2,i},-R_{3,i})^T \; ,
\end{eqnarray}
and $\lambda \simeq 0.1$. The subscript ``$i$'' denotes that the expressions are evaluated at $(r_i,L_{z,i},E_i)$. We stop iterating when $|B_i| < x$, where $x \simeq 10^{-7}$. The method outlined here works well for a large fraction of parameter space.

\section{Numerical integration across turning points}
\label{app:tp}
As mentioned in Sec.\ \ref{sec:ecc_pres}, Eq.\ (\ref{eq:accn3}) passes through turning points. The numerical integrator can accumulate error when $dr/dt \rightarrow 0$. This section describes our algorithm to resolve the issue. 

Let $t_p$ denote the instant at which $dr/dt=0$. The radial motion is highly symmetric about the turning point. Thus, we must have,
\begin{eqnarray}
\label{eq:symm}
\left. \frac{dr}{dt}\right|_{t_p+\epsilon} & = & -\left. \frac{dr}{dt}\right|_{t_p-\epsilon} \;,
\end{eqnarray}
where $\epsilon$ is an infinitesimal duration of time. When the radial velocity becomes very small, we exploit this symmetry and set 
\begin{eqnarray}
\left. \frac{dr}{dt}\right|_{t_p+\delta t} = - \left. \frac{dr}{dt}\right|_{t_p-\delta t}\;,
\end{eqnarray}
which is the discretized version of Eq.\ (\ref{eq:symm}).

\end{document}